\documentclass[a4paper]{article}
\usepackage{amssymb}
\usepackage{amsmath}

\catcode`\@=11

\@addtoreset{equation}{section}
\catcode`\@=12

\DeclareMathOperator{\C}{\mathbb C}
\DeclareMathOperator*{\res}{\rm res}
\DeclareMathOperator{\sgn}{\rm sgn}
\DeclareMathOperator{\arccot}{\rm arccot}

\begin{document}

\title{Inverse scattering theory for the perturbed 1-soliton potential of
the heat equation \thanks{Work supported in part by INTAS 99-1782, by Russian Foundation for
Basic Research 99-01-00151 and 00-15-96046 and by COFIN 2000 ``Sintesi''.}}
\date{June 19, 2001}
\author{M. Boiti, F. Pempinelli, A.K. Pogrebkov${}^\dag$ and B. Prinari \\
{\ }\\
Dipartimento di Fisica dell'Universit\`{a}
and Sezione INFN, 73100 Lecce, Italy\\
${}^\dag$Steklov Mathematical Institute
Moscow, 117966, GSP-1, Russia}
\maketitle

\begin{abstract}
Inverse scattering transform method of the heat equation is developed for a special subclass of
potentials nondecaying at space infinity---perturbations of the one-soliton potential by means
of decaying two-dimensional functions. Extended resolvent, Green's functions, and Jost
solutions are introduced and their properties are investigated in detail. The singularity
structure of the spectral data is given and then the Inverse problem is formulated in an
exact distributional sense.
\end{abstract}

\section{Introduction}

The equation of the heat conduction, or heat equation for short,
 \begin{equation}
 \mathcal{L}\Phi(x)=0,\label{heat}
 \end{equation}
where the operator
 \begin{equation}
 \mathcal{L}(x,\partial _{x})=-\partial _{x_{2}}+\partial _{x_{1}}^{2}-u(x),\qquad
 x=(x_{1},x_{2})  \label{1}
 \end{equation}
for more than 25 years has been known~\cite{dryuma74,zakharov74} to be associated to the
Kadomtsev--Petviashvili (more precisely,  KPII) equation
 \begin{equation}
 (u_{t}-6uu_{x_{1}}+u_{x_{1}x_{1}x_{1}})_{x_{1}}=-3u_{x_{2}x_{2}}. \label{KPII}
 \end{equation}
The scattering theory for the equation of heat conduction with a real potential $u(x)$ was
developed in \cite{BarYacoov}--\cite{Grinevich}, but only the case of potentials rapidly
decaying at large distances on the $x$-plane was considered. On the other side, it is well
known that~(\ref{KPII}) is a (2+1)-dimensional generalization of the famous KdV equation: if
the function $u_{1}(t,x_{1})$ obeys KdV then
 \begin{equation}
 u(t,x_{1},x_{2})=u_{1}(t,x_{1}+\mu x_{2}-3\mu ^{2}t)  \label{1to2}
 \end{equation}
solves~(\ref{KPII}) for an arbitrary constant $\mu \in \mathbb{R}$. Thus it is natural to
consider solutions of~(\ref{KPII}) that are not decaying in all directions at space infinity
but have 1-dimensional rays with behavior of the type~(\ref{1to2}). The scattering theory for
the operator~(\ref{1}) with such potentials is absent in the literature. Moreover, it is easy to
observe that like in the KPI case (see~\cite{total}) the standard integral equation for the
Jost solution~\cite{BarYacoov} is meaningless for this situation and does not determine the
solution itself. In trying to solve this problem for the nonstationary Schr\"{o}dinger
operator, associated to the KPI equation, a new general approach to the inverse scattering
theory was introduced, which was called resolvent approach, see \cite{total}--\cite{PhysicaD}
and references therein. In \cite{KPII} we developed the scattering theory for the $N$-soliton
solutions given in terms of B\"{a}cklund transformations of the decaying background potential.
These results for the simplest case $N=1$ are essentially used below.

Here we apply the resolvent approach to the heat equation~(\ref{1}) with a potential $u(x)$
that is a perturbation of a 1-dimensional potential $u_{1}(x)$ of the kind~(\ref{1to2}) by
means of a potential $u_{2}(x)$ rapidly decaying in all directions
 \begin{equation}
 u(x)=u_{1}(x)+u_{2}(x).  \label{potential}
 \end{equation}
We introduce and study properties of the resolvent, dressing operators, Jost solutions and
scattering data and formulate the inverse problem relevant to this case. In fact we consider
here the  simplified version of~(\ref{1to2}) in which $\mu =0$. The generic case is
reconstructed by means of the Galilean invariance of~(\ref{KPII}). Thus in what follows
$u_{1}(x)\equiv u_{1}(x_{1})$ and, moreover, we consider for simplicity the case where $u_{1}$
is the 1-dimensional soliton potential (see~(\ref{1-5}) below).

Thus here we apply the Inverse scattering theory to a non-scattering situation since the
``obstacle" is infinite. Such extension of the Inverse scattering theory results in the new
and unexpected properties of familiar objects, like the Jost solutions and the spectral data.
We show that they get specific singularities in the complex domain of the spectral parameter.
Derivation and description of these singularities are our main results here. The article is
organized as follows. In Sec.~2 we sketch some general aspects of the resolvent approach that
are necessary for our construction. In Sec.~3 we present results of embedding the theory of
the one-dimensional one-soliton potential in two dimensions. Presentation here is based on the
work~\cite{KPII}. We describe in detail properties of the extended resolvent and Green's
functions of operator~(\ref{1}) with $u(x)=u_1(x_1)$. On this basis in Sec.~4 the resolvent of
the operator~(\ref{1}) now with the generic potential $u(x)$ given in~(\ref{potential}) is
introduced and its properties are described. The departure from analyticity of the resolvent
leads us to definitions of the Jost solutions and spectral data and description of their
properties (Sec.~5). In this way we supply all terms of the Inverse problem with proper
meaning in terms of distributions. In Conclusion some generalizations and future developments
of these results are discussed. Main results of this article were announced in our
work~\cite{art1}.

\section{Extension of differential operators and resolvent}

In the framework of the resolvent approach we work in the space $\mathcal{S}'$ of tempered
distributions $A(x,x';q)$ of the six real variables $x=(x_1,x_2),x',q\in \mathbb{R}^{2}$. It is
convenient to consider $q$ as the imaginary part of a two-dimensional complex variable
 $\mathbf{q=q}_{\Re }+i\mathbf{q}_{\Im }=(\mathbf{q}_{1},\mathbf{q}_{2})\in \mathbb{C}^{2}$
and to introduce the ``shifted'' Fourier transform
\begin{equation}
A(p;\mathbf{q})=\frac{1}{(2\pi )^{2}}\int dx\int dx'\,e^{i(p+
\mathbf{q}_{\Re })x-i\mathbf{q}_{\Re }x'}A(x,x';\mathbf{q}_{\Im })  \label{2a}
\end{equation}
where $p\in \mathbb{R}^{2}$, $px=p_{1}x_{1}+p_{2}x_{2}$ and
 $\mathbf{q}_{\Re }x= \mathbf{q}_{1\Re }x_{1}+\mathbf{q}_{2\Re }x_{2}$.
We consider the distributions $A(x,x';q)$ and $A(p;\mathbf{q})$ as kernels in two different representations,
the $x$- and $p$-representation, respectively, of the operator $A(q)$ ($A$ for short). The
composition law in the $x$-representation is defined in the standard way, that is
 \begin{equation}
 (AB)(x,x';q)=\int dx''\,A(x,x'';q)\,B(x'',x';q). \label{3}
 \end{equation}
Since the kernels are distributions this composition is neither necessarily defined for all
pairs of operators nor associative. In terms of the $p$-representation~(\ref{2a}) this
composition law is given by a sort of a ``shifted'' convolution,
 $(AB)(p;\mathbf{q})=\int dp'A(p-p';\mathbf{q}+p')B(p';\mathbf{q})$.
On the space of these operators we define the conjugation $A^{\ast }$, that in the
$x$-representations reads as
 \begin{equation}
 A_{}^{\ast }(x,x';q)=\overline{A(x,x';q)}, \label{3a}
 \end{equation}
where bar denotes complex conjugation,  or as
 $A^{\ast}(p;\mathbf{q})=\overline{A(-p;-\overline{\mathbf{q}})}$
in the $p$-representation. Below we say that the operator $A(q)$ is real if
 $A_{}^{\ast}(q)=A(q)$,
that in terms of $p$-representation means that
 $\overline{A(p;\mathbf{q})}=A(-p;-\overline{\mathbf{q}})$.
The set of differential operators
 $\mathcal{D}(x,\partial _{x})=\sum d_{n}(x)\partial _{x}^{n}$ is embedded in the introduced
space of operators by considering the operators $D$ with kernel
 $D(x,x')=\mathcal{D}(x,\partial _{x})\delta (x-x')$, where
 $\delta (x)=\delta (x_{1})\delta (x_{2})$ is the two-dimensional $\delta $-function and,
 then, by mapping them in the operators $D(q)$ with kernel
 \begin{equation}
 D(x,x';q)\equiv e^{-q(x-x')}D(x,x')=\mathcal{D}
 (x,\partial _{x}+q)\delta (x-x'),  \label{5}
 \end{equation}
to which we refer as the \textbf{extended} version of the differential operator $\mathcal{D}$.
The notion of reality for a differential operator $D$ is exactly the condition that its
coefficients $d_{n}(x)$ are real.

For the operator~(\ref{1}) the extension $L(q)$ is given by
 \begin{equation}
 L=L_{0}-U,  \label{6}
 \end{equation}
where $L_{0}$ is the extension of $\mathcal{L}(x,\partial _{x})$ in the case of zero potential,
i.e.\ it has kernels
 \begin{equation}
 L_{0}(x,x';q)=\left[ -(\partial _{x_{2}}+q_{2})+(\partial_{x_{1}}+q_{1})^{2}\right]
 \delta (x-x'),\qquad L_{0}(p;\mathbf{q})=(i\mathbf{q}_{2}-\mathbf{q}_{1}^{2})\delta (p),
 \label{7}
 \end{equation}
and the multiplication operator $U$ can be called the potential since it has kernel
 \begin{equation}
 U(x,x';q)=u(x)\delta (x-x'). \label{8}
 \end{equation}
Below we always suppose that $u(x)$ is real, which by~(\ref{3a}) means that the
operator~(\ref{1}) is real also: $L^*=L$.

The main object of our approach is the \textbf{extended resolvent} $M(q)$ of the operator
$L(q)$, which is defined as the inverse of the operator $L$, that is
\begin{equation}
LM=ML=I,  \label{10}
\end{equation}
in the space of operators. Here $I$ is the unity operator, $I(x,x';q)=\delta (x-x')$,
$I(p;\mathbf{q})=\delta (p)$. In order to make this inversion uniquely defined we impose the
condition that the product
\begin{equation}
\int dp'M(p-p';\mathbf{q}+\mathbf{s}+p')M(p';\mathbf{q}) \label{condition}
\end{equation}
exists as distribution in $p$ and $\mathbf{q}$ and that it is a continuous function of
$\mathbf{s}$ in a neighborhood of $\mathbf{s}=0$ when $\mathbf{s}\neq 0$.

Thanks to definitions~(\ref{6}),~(\ref{7}), and~(\ref{10}) $M$ is real and in particular, the
resolvent $M_{0}$ of the bare operator $L_{0}$ has in the $p$-representation kernel
 $M_{0}(p;\mathbf{q})=\delta (p)(i\mathbf{q}_{2}-\mathbf{q}_{1}^{2})^{-1}$.
As function of $\mathbf{q}$ it is singular when $\mathbf{q}=\ell(\mathbf{q}_{1})$, where the
special two-component vector
 \begin{equation}
 \ell (k)=(k,-ik^{2})  \label{l}
 \end{equation}
was introduced. In the $x$-representation by inverting~(\ref{2a}) we get
 \begin{equation}
 M_{0}^{}(x,x';q)=\frac{1}{2\pi }\int d\alpha \,\bigl[\theta
 (q_{1}^{2}-\alpha _{}^{2}-q_{2}^{})-\theta (x_{2}^{}-x_{2}')
 \bigr]\,e_{}^{-(i\ell (\alpha +iq_{1})+q)(x-x')}.  \label{M0}
 \end{equation}

For a generic operator $A$ with kernel $A(x,x';q)$ the operation inverse to the extension
procedure, defined in~(\ref{5}) for a differential operator, is given by
 \begin{equation}
 \widehat{A}(x,x';q)=e^{q(x-x')}A(x,x';q). \label{11}
 \end{equation}
In contrast with the case of the extended differential operators for which
 $\widehat{D}(x,x';q)=D(x,x')\equiv \mathcal{D}(x\mathcal{,}
 \partial _{x})\delta (x-x')$, in general $\widehat{A}(x,x^{\prime};q)$
does depend on $q$ and, moreover, can have an exponential growth at space infinity. Therefore
$\widehat{A}(x,x';q)$ not necessarily belongs to the space $\mathcal{S}'$ of tempered
distributions. The fact that $\widehat{A}(x,x';q)$ can depend on $q$ will play a crucial role
in the following. For instance also in the case of the simplest resolvent~(\ref{M0}) we have
that the function $\widehat{M}_{0}(x,x';q)$ depends effectively on the variable $q$ and is
exponentially growing at space infinity. More generally from~(\ref{10}) we have
 \begin{equation}
 \mathcal{L}(x,\partial _{x})\widehat{M}(x,x';q)=\mathcal{L}^{\text{d}}(x',\partial _{x'}^{})
 \widehat{M}(x,x';q)=\delta (x-x'),
 \label{12}
 \end{equation}
where $\mathcal{L}^{\text{d}}$ is the operator dual to $\mathcal{L}$. The function
$\widehat{M}(x,x';q)$ can be considered a parametric ($q\in \mathbb{R}^{2}$) family of Green's
functions of the operator $\mathcal{L}$. In what follows we use special notations for the
equalities of the type~(\ref{12}), writing them as
\begin{equation}
\overrightarrow{\mathcal{L}}\widehat{M}(q)=\widehat{M}(q)\overleftarrow{\mathcal{L}}=I,
\label{13}
\end{equation}
where $\overrightarrow{\mathcal{L}}$ denotes the operator $\mathcal{L}$ applied to the
$x$-variable of the function $\widehat{M}(x,x';q)$ and $\overleftarrow{\mathcal{L}}$ denotes
the operator dual to $\mathcal{L}$ applied to the $x'$-variable of the same function. Operation
(\ref{11}) has no analog in terms of the $p$-representation. Nevertheless, local properties of
the kernels in the $x$-representation are preserved, and we use the kernels with the hat in
what follows intensively.

Thanks to our definitions~(\ref{2a}) and~(\ref{5}) it is easy to see that in terms of the
$p$-representation the dependence on the $\mathbf{q}$-variables of the kernels of the extension
of a differential operator is polynomial (like in the example~(\ref{7})). Correspondingly, the
essential role in the study of the properties of the resolvent is played by the investigation
of its departure from analyticity, in particular, by its d-bar derivatives with respect to the
$\mathbf{q}$-variables. Thus to a generic operator $A$ with kernel $A(p;\mathbf{q})$ in the
$p$-representation we associate two operators $\bar{\partial}_{j}A$ with kernels
 \begin{equation}
 (\bar{\partial}_{j}^{}A)(p;\mathbf{q})=\frac{\partial A(p;\mathbf{q})}
 {\partial \overline{\mathbf{q}}_{j}^{}},\quad j=1,2,  \label{151}
 \end{equation}
where the derivatives are considered in the sense of distributions. In terms of the objects
introduced in~(\ref{11}) we get by inversion of~(\ref{2a}) that
\begin{equation}
(\widehat{\bar{\partial}_{j}A})(q)=\frac{i}{2}\frac{\partial \widehat{A}(q)}{
\partial q_{j}}.  \label{16}
\end{equation}

Multiplying equalities in~(\ref{10}) from the left and right, correspondingly, by $M_{0}$ we
get thanks to~(\ref{6})
 \begin{equation}
 M=M_{0}^{}+M_{0}^{}UM,\qquad M=M_{0}^{}+MUM_{0}^{}. \label{integralM}
 \end{equation}
Since the resolvent $M_{0}$ is explicitly given these are integral
equations determining the solution $M$ of~(\ref{10}).
In the literature (see, say, \cite{Wickerhauser} and \cite {Grinevich})
on the Jost solutions of the heat equation some small norm conditions on the potential $u$ are
known to guarantee the existence of the Jost solutions. So it is natural to assume that under
such conditions the solution $M$ of the above integral equations exists and is unique (the same
for both integral equations). In this case the resolvent $M$ can be considered as a small
perturbation of the resolvent $M_{0}$ and this bare resolvent determines the properties of $M$
by means of~(\ref{integralM}). The main problem of construction of the Inverse scattering
transform for the operator~(\ref{1}) is that the potential $u(x)$ in~(\ref{potential}) does not
obey any small norm condition.

In order to overcome this difficulty we use a so called Inverse scattering transform on
a non trivial background~\cite{PhysicaD}. Let us consider a kind of Hilbert identity, known in the
standard spectral theory of operators. Precisely, if $M(q)$ is the extended resolvent of the
operator $L(q)$ with potential $u$ and $M'(q)$ the extended resolvent of the operator $L'(q)$
with a different potential $u'$, then, by~(\ref{10}) we have
\begin{equation}
M'-M=-M'(L'-L)M.  \label{H1}
\end{equation}
Strictly speaking, this follows under the assumption that the product in the r.h.s.\ is
associative. This is a natural assumption since $L'(x,x';q)-L(x,x';q)=(u(x)-u(x'))\delta (x-x')$
and $M$ satisfies condition~(\ref{condition}). Let now $ \mathcal{L}_{1}$ denote the operator
(\ref{1}) in the special case where the potential $u(x)$ in~(\ref{potential}) is purely
1-dimensional, i.e.\ $u_{2}(x)\equiv 0$. Let $L_{1}$ denote its extension and $M_{1}$ its
resolvent, that is let (cf.~(\ref{6}))
 \begin{align}
 L_{1}^{}& =L_{0}^{}-U_{1}^{},\qquad \mathcal{L}_{1}^{}(x,\partial _{x}^{}) =-\partial
 _{x_{2}}^{}+\partial _{x_{1}}^{2}-u_{1}^{}(x),  \label{1-1} \\
 L& =L_{1}^{}-U_{2}^{}, \qquad L_{1}^{}M_{1}^{} =M_{1}^{}L_{1}^{}=I,  \label{1-3}
 \end{align}
where as in~(\ref{8}) $U_j(x,x';q)=u_j(x)\delta(x-x')$. Choosing now in~(\ref{H1}) $L'=L_{1}$
we get
 \begin{equation}
 M=M_{1}^{}+M_{1}^{}U_{2}^{}M,\quad M=M_{1}^{}+MU_{2}^{}M_{1}^{}, \label{17}
 \end{equation}
where the second equality is derived in analogy. These equations generalize~(\ref{integralM})
for the case where $M_1$ is non trivial and if the resolvent $M_{1}$ is known they also can be
considered as defining the resolvent $M$. If we choose $U_{2}$ obeying the small norm condition
mentioned above, we can assume that the solution $M$ of both equations~(\ref{17}) exists and
is unique. Then, thanks to~(\ref{1-3}) $M$ obeys~(\ref{10}). Contrary to~(\ref{integralM}) now
$M$ can be considered a perturbation of the resolvent $M_{1}$. So in the next section we study
the properties of the resolvent $M_1$ in detail.

\section{One-dimensional potential}

We already mentioned in the Introduction that in this article we deal with the case where
$u_{1}$ in~(\ref{potential}) is the 1-dimensional soliton potential
\begin{equation}
u_{1}^{}(x)=\frac{-2a^{2}}{\cosh ^{2}\left[ a\left( x_{1}^{}-x_{0}^{}\right) \right] }\,,
\label{1-5}
\end{equation}
with $a>0$ and $x_{0}$ real constants. In this section we consider the case where the
perturbation is absent, $u_{2}\equiv 0$. We re-formulate in the 2-dimensional space the well
known facts about Jost solutions of this 1-soliton potential and introduce and study the
properties of the resolvent and Green's functions in this case.

The differential equations $\mathcal{L}_{1}(x,\partial _{x})\Phi _{1}(x,k)=0$,
$\mathcal{L}_{1}^{\text{d}}(x,\partial _{x})\Psi _{1}(x,k)=0$ for the Jost solution $\Phi
_{1}(x,k)$ and its dual $\Psi _{1}(x,k)$, by using the notation introduced in~(\ref{13}), can
be re-written as follows
\begin{equation}
\overrightarrow{\mathcal{L}}_{1}^{}\Phi _{1}^{}(k)=0,\quad \Psi
_{1}^{}(k)\overleftarrow{\mathcal{L}}_{1}^{}=0.  \label{1-6}
\end{equation}
Here and in the following we omit the $x$-dependence when it is irrelevant.
These solutions are given explicitly by
\begin{align}
\Phi _{1}^{}(x,k)& =\frac{k-ia\tanh \left[ a(x_{1}^{}-x_{0}^{})\right]
}{k-ia}\,e_{}^{-i\ell (k)x},  \label{1-7} \\
\Psi _{1}^{}(x,k)& =\frac{k+ia\tanh \left[ a(x_{1}^{}-x_{0}^{})\right] }{k+ia}\,e_{}^{i\ell
(k)x},  \label{1-8}
\end{align}
where $k\in \C$ and the two-component vector $\ell (k)$ is defined in~(\ref{l}).
They obey the conjugation properties
\begin{equation}
\overline{\Phi _{1}^{}(x,k)}=\Phi _{1}^{}(x,-\bar{k}),\quad \overline{
\Psi _{1}^{}(x,k)}=\Psi _{1}^{}(x,-\bar{k})  \label{1-9}
\end{equation}
that are equivalent to the reality condition for the potential $u_{1}$, and are normalized at
$k$-infinity as follows
\begin{equation}
\lim_{k\rightarrow \infty }e_{}^{i\ell (k)x}\Phi _{1}^{}(x,k)=1,\qquad \lim_{k\rightarrow
\infty }e_{}^{-i\ell (k)x}\Psi _{1}^{}(x,k)=1. \label{1-10}
\end{equation}

The functions $\Phi _{1}(x,k)$ and $\Psi _{1}(x,k)$ are meromorphic in the
complex domain of the spectral parameter $k$ with poles at $k=ia$ and $k=-ia$,
correspondingly. Thus, these functions obey the d-bar equations
\begin{equation}
\frac{\partial \Phi _{1}^{}(x,k)}{\partial \bar{k}}=i\pi \Phi
_{1,a}^{}(x)\delta (k-ia),\qquad \frac{\partial \Psi _{1}^{}(x,k)}{
\partial \bar{k}}=i\pi \Psi _{1,-a}^{}(x)\delta (k+ia),  \label{1-11}
\end{equation}
where we introduced the notations
\begin{equation}
\Phi _{1,a}^{}(x)=-i\res\limits_{k=ia}\Phi _{1}^{}(x,k),\qquad \Psi
_{1,-a}^{}(x)=-i\res\limits_{k=-ia}\Psi _{1}^{}(x,k).  \label{1-12}
\end{equation}
Explicitly we have
\begin{equation}
\Phi _{1,a}(x)=\frac{ae^{ax_{0}+a^{2}x_{2}}}{\cosh \left[
a(x_{1}^{}-x_{0}^{})\right] },\quad \Psi _{1,-a}(x)=-\frac{
ae^{ax_{0}-a^{2}x_{2}}}{\cosh \left[ a(x_{1}^{}-x_{0}^{})\right] }. \label{1-13}
\end{equation}
Let
\begin{equation}
c=2ae^{2ax_{0}}
\end{equation}
and $\Phi _{1,-a}(x)$ and $\Psi _{1,a}(x)$ the values of the Jost
solutions in the conjugated points,
\begin{equation}
\Phi _{1,-a}^{}(x)=\Phi _{1}^{}(x,-ia),\qquad \Psi _{1,a}^{}(x)=\Psi _{1}^{}(x,ia).
\label{1-15}
\end{equation}
Then thanks to~(\ref{1-7}),~(\ref{1-8}), and~(\ref{1-13}) the Jost solutions obey in the
complex domain of the spectral parameter the following scalar products
 \begin{align}
 \int dx_{1}\Psi _{1}^{}(x,k+p)\Phi _{1}^{}(x,k)& =2\pi \delta (p),\quad
 p\in \mathbb{R},  \label{1-16} \\
 c\int dx_{1}^{}\,\Phi _{1,-a}^{}(x)\Psi _{1,a}^{}(x)& =1,  \label{1-17}\\
 \int dx_{1}^{}\,\Psi _{1,a}^{}(x)\Phi _{1}^{}(x,k)& =0,\quad k_{\Im}^{2}<a_{}^{2},
 \label{1-18} \\
 \int dx_{1}^{}\,\Phi _{1,-a}^{}(x)\Psi _{1}^{}(x,k)& =0,\quad k_{\Im }^{2}<a_{}^{2},
 \label{1-19}
 \end{align}
and the completeness relation
 \begin{equation}
 \frac{1}{2\pi }\int\limits_{x_{2}=x_{2}'}dk_{\Re }\,\Phi _{1}^{}(x,k)\Psi _{1}^{}(x',k)+c\theta
 (a_{}^{2}-k_{\Im }^{2})\Phi _{1,-a}^{}(x)\Psi _{1,a}^{}(x')\Bigl|_{x_{2}=x_{2}'}=\delta
 (x_{1}^{}-x_{1}'). \label{1-20}
 \end{equation}

Equations~(\ref{1-11}) can be considered as two Inverse problems defining the Jost solution and
the dual Jost solution. The formulation of these problems is closed by giving the normalization
conditions~(\ref{1-10}) and the following relations:
\begin{equation}
\Phi _{1,a}^{}(x)=c\Phi _{1,-a}^{}(x),\qquad \Psi _{1,-a}^{}(x)=-c\Psi _{1,a}^{}(x),
\label{1-14}
\end{equation}
where $\Phi _{1,-a}(x)$ and $\Psi _{1,a}(x)$ are defined in~(\ref{1-15}).

These formulae show that the embedding in two dimensions of the Jost solutions of the 1-soliton
potential is trivial and just mimics the 1-dimensional construction up to the warning that, due
to their exponential dependence on $x_{2}$, the functions $\Phi _{1,a}(x)$ and $\Psi
_{1,-a}(x)$ are not square integrable with respect to the $x$-variables and, therefore, are not
eigenfunctions of the operator $\mathcal{L}_{1}$.

On the contrary, the 2-dimensional resolvent $M_{1}(q)$ as well as the Green's function $G_{1}$
of the 2-dimensional operator $\mathcal{L}_{1}$ with the 1-dimensional potential $u_{1}$ are
not trivial extensions of the
corresponding 1-dimensional objects associated to the operator $
(k^{2}+\partial _{x_{1}}^{2}-u_{1}(x_{1}))$. In terms of the Jost solutions
introduced above we can write the kernel of this resolvent obtained in \cite{KPII} as
\begin{align}
\widehat{M}_{1}^{}(x,x';q)& =\frac{1}{2\pi }\int\limits_{k_{\Im }=q_{1}}dk_{\Re
}^{}\,\bigl[\theta (q_{1}^{2}-q_{2}^{}-k_{\Re }^{2})-\theta (x_{2}^{}-x_{2}')\bigr]\,\Phi
_{1}^{}(x,k)\Psi
_{1}^{}(x',k)+  \notag \\
& +c\theta (a_{}^{2}-q_{1}^{2})\bigl[\theta (a_{}^{2}-q_{2}^{})-\theta
(x_{2}^{}-x_{2}')\bigr]\Phi _{1,-a}^{}(x)\Psi _{1,a}^{}(x') \label{1-21}
\end{align}
where the hat over the kernel is used in the sense of notation~(\ref{11}).

Thanks to the equalities~(\ref{1-6}) and~(\ref{1-20}) it is easy to check directly that
$\widehat{M}_{1}(q)$ obeys the equations
\begin{equation}
\overrightarrow{\mathcal{L}}_{1}^{}\widehat{M}_{1}^{}(q)=\widehat{M}_{1}^{}(q)
\overleftarrow{\mathcal{L}}_{1}^{}=I,  \label{1-22}
\end{equation}
that means (cf.~(\ref{13})) that $M_{1}(q)$ obeys~(\ref{1-3}) and is indeed the inverse of the
operator $L_{1}(q)$. Moreover, using the explicit
formulas~(\ref{1-7}),~(\ref{1-8}), and~(\ref{1-13}) we get that $
M_{1}(x,x';q)\in \mathcal{S}'(R^{6})$, i.e.\ it belongs to the space of operators under
consideration. It can also be proved directly that $M_{1}$ obeys condition~(\ref{condition}),
so it is the extended resolvent according to our definition. By means of~(\ref{1-9}) we get
also that $M_{1}$ is real, $M_{1}^*=M_{1}$ according to definition~(\ref{3a}).

We emphasize that in order to prove these results it is not necessary to use the explicit
formulae for $\Phi _{1}$ and $\Psi _{1}$ but only their general properties. In fact, if one
considers a 1-dimensional potential $u_{1}$ which, in addition, has a non trivial continuous
spectrum one gets the same formula for the resolvent $M_{1}$. If the discrete (1-dimensional)
spectrum
of $u_{1}$ contains $N$ solitons with parameters $a_{j}$ and $c_{j}$ ($
j=1,2,\dots ,N$), then the last term in~(\ref{1-21}) must be substituted by the sum of similar
terms each corresponding to a value of $j$.

Now we describe in detail the properties of $M_{1}(x.x';q)$. The first term in the r.h.s.\ of
(\ref{1-21}) is a continuous function of $q=(q_1,q_2)$ on the $q$-plane with discontinuities
on the lines $q_{1}=\pm a$ due to the pole singularities of $\Phi _{1}(k)$ and $\Psi _{1}(k)$.
The second term thanks to the $\theta $ functions has discontinuities on the lines $q_{1}=\pm
a$ and on the cut $q_{2}=a^{2}$, $|q_{1}|<a$. The singularities on the lines $q_{1}=\pm a$ are
exactly compensated among the two terms. Thus the kernel $M_{1}(x,x';q)$ is a continuous
function of $q$ with a discontinuity on the cut $q_{2}=a^{2}$, $\left| q_{1}\right| <a$. This
discontinuity is specific of the potential $u_{1}(x)$, or more generally of
a potential with discrete spectrum and it gives the essential difference of $
M_{1}$ with respect to the bare resolvent $M_{0}$~(\ref{M0}). Let us underline that, in spite
of the fact that $\mathcal{L}_{1}$ applied to the term with $\theta (a^{2}-q_{2})$ that causes
this discontinuity gives zero, this term cannot be omitted in~(\ref{1-21}), since only thanks
to the fact that $\theta (x_{2}-x_{2}')$ and $\theta (a^{2}-q_{2})$ have
opposite signs the kernel $M_{1}(x,x';q)\equiv e^{-q(x-x')}
\widehat{M}_{1}(x,x';q)$ is a tempered distribution with respect to the $x$-variables.

The kernel $M_{1}(p;\mathbf{q})$ in the $p$-representation is not an analytic function of
$\mathbf{q}$. By~(\ref{16}) the d-bar derivatives of $M_1$ with respect to $\mathbf{q}_{j}$ are
proportional to $\partial \widehat{M}_1/\partial q_j$ and for the latter we get
from~(\ref{1-21}) equalities
\begin{align}
\frac{\partial \widehat{M}_{1}^{}(q)}{\partial q_{1}^{}}=& \frac{i}{\pi }
\int\limits_{k_{\Im }=q_{1}}dk_{\Re }^{}\,\bar{k}\delta (\ell _{2\Im }^{}(k)-q_{2}^{})\,\Phi
_{1}^{}(k)\otimes \Psi _{1}^{}(k),
\label{1-44} \\
\frac{\partial \widehat{M}_{1}^{}(q)}{\partial q_{2}^{}}=& \frac{-1}{
2\pi }\int\limits_{k_{\Im }=q_{1}}dk_{\Re }^{}\,\delta (\ell _{2\Im }^{}(k)-q_{2}^{})\,\Phi
_{1}^{}(k)\otimes \Psi _{1}^{}(k), \label{1-45}
\end{align}
where $(\Phi _{1}(k)\otimes \Psi _{1}(k))(x,x')\equiv \Phi _{1}(x,k)\Psi _{1}(x',k)$ is the
standard direct product and where by the above discussion we consider $q_{2}\neq a^{2}$. For
the discontinuity along this line we get
\begin{equation}
\widehat{M}_{1}^{}(q)\Bigl|_{q_{2}=a_{}^{2}+0}^{}-\widehat{M}_{1}^{}(q)
\Bigl|_{q_{2}=a_{}^{2}-0}^{}=-c\theta (a_{}^{2}-q_{1}^{2})\Phi _{1,-a}^{}\otimes \Psi
_{1,a}^{}.  \label{1-46}
\end{equation}

We have to study now the behavior of $M_{1}(q)$ at the end points of the cut, i.e.\ when
 $q\sim(\pm a,a^{2})$. Firstly, it is convenient to subtract to $M_{1}(q)$ its value, say, on
the upper or lower edges of the cut:
 \begin{equation}
 g_{1}^{\pm}=\lim_{q_{2}=a^{2}\pm 0}\widehat{M}_{1}^{}(q)\Bigl|_{|q_{1}|<a}^{}.  \label{1-23}
 \end{equation}
Since $\Phi _{1}^{}(k)$ and $\Psi _{1}^{}(k)$ are analytic for $|k_{\Im }|<a$, we deduce
from~(\ref{1-21}) that $g_{1}^{\pm}$ are independent also of $q_{1}$ and their kernels equal
 \begin{equation}
 g_{1}^{\pm}(x,x')=-\frac{\theta (x_{2}^{}-x_{2}')}{2\pi }\int d\alpha \,\Phi
 _{1}^{}(x,\alpha )\Psi _{1}^{}(x',\alpha )\mp c\theta (\pm(x_{2}^{}-x_{2}'))
 \Phi _{1,-a}^{}(x)\Psi_{1,a}^{}(x'),  \label{M1rega}
 \end{equation}
where $\int d\alpha $ denotes integration along the whole real axis. Now extracting explicitly
from the first term in the r.h.s.\ of~(\ref{1-21}) the contribution coming from the poles of
$\Phi _{1}^{}(k)$ and $\Psi _{1}^{}(k)$ we get that, say, difference
 $\widehat{M}_{1}^{}(q)-g_{1}^{-}$ behaves at points $q=(\pm a,a^{2})$ as
 \begin{multline}
 \widehat{M}_{1}^{}(q)-g_{1}^{-}=-c\left( \frac{\theta (q_{1}^{2}-q_{2}^{})}{\pi
 }\arccot\frac{a-|q_{1}^{}|}{\sqrt{q_{1}^{2}-q_{2}^{}}}+\theta (q_{2}^{}-q_{1}^{2})\theta
 (q_{2}^{}-a_{}^{2})\right) \Phi _{1,-a}^{}\otimes \Psi _{1,a}^{}+ \\
 +o(1),\quad q\sim (\pm a,a_{}^{2}).  \label{1-25}
 \end{multline}
Thus $\widehat{M}_{1}(q)$ is bounded but discontinuous at $q=(\pm a,a^{2})$, while its regular
part, $g_{1}^{-}$, is the same for both these points.

Now it is easy to see that the discontinuity of the resolvent along the cut $q_{2}=a^{2}$,
$|q_{1}| <a$ and the ill definiteness at the points $q=(\pm a,a^{2})$ are the result of
embedding the 1-dimensional potential in the two dimensional space. Indeed, the resolvent of
the Sturm--Liouville operator $\partial _{x_{1}}^{2}-u_{1}(x_{1})-q_{2}$ is obtained from
$M_{1}(q)$ by means of the operation
 $\int dx_{2}\,e^{-q_{2}(x_{2}-x_{2}')}\widehat{M}_{1}(x,x';q)$. By~(\ref{1-21})
and~(\ref{1-7}),~(\ref{1-8}),~(\ref{1-13}) we get the standard expression for the 1-dimensional
Green's function with a pole at $q_{2}=a^{2}$.

We already noted that $\widehat{M}_{1}^{}(x,x';q)$ defines a family of Green's functions. Among
them we expect should play a special role those obtained considering the values of $q$
belonging to the support of the defects of analyticity given in~(\ref{1-44}),~(\ref{1-45}) and
in~(\ref{1-46}). We consider, therefore, the Green's functions
 \begin{align}
 G_{1}^{}(x,x',k)& =\widehat{M}_{1}^{}(x,x';q)\Bigr|_{q=\ell _{\Im }^{}(k)}^{},  \label{1-47} \\
 G_{1}^{\pm }(x,x';k)& =\widehat{M}_{1}^{}(x,x';q)\Bigr|_{q_{1}=k_{\Im},\,q_{2}=a^{2}\pm 0}^{},
 \label{1-48}
 \end{align}
where $k\in \mathbb{C}$ is the spectral parameter and we denote $
q_{1}=k_{\Im }$ (see~(\ref{l})) in order to meet the standard notation. From these definitions
it follows directly that
 \begin{align}
 & \overrightarrow{\mathcal{L}}_{1}^{}G_{1}^{}(k)=G_{1}^{}(k)
 \overleftarrow{\mathcal{L}}_{1}^{}=I,\qquad \overrightarrow{\mathcal{L}}
 _{1}^{}G_{1}^{\pm }(k)=G_{1}^{\pm }(k)\overleftarrow{\mathcal{L}}
 _{1}^{}=I,  \label{1-50} \\
 & \overline{G_{1}^{}(k)}\equiv G_{1}^{}(-\bar{k})\equiv G_{1}^{}(k),
 \label{1-51} \\
 & G_{1}^{}(k)\Bigl|_{k_{\Re }=0}^{}=G_{1}^{}(0),\qquad G_{1}^{\pm }(k)=G_{1}^{\pm }(ik_{\Im
 }^{}),  \label{1-52}
 \end{align}
i.e.\ $G_{1}^{\pm }(k)$ are independent on $k_\Re$ and then inside the strip they coincide with
$g_1^\pm$ introduced in~(\ref{1-23}),
 \begin{equation}
 G_{1}^{\pm }(x,x';k)\Bigr|_{|k_{\Im}|<a}^{} =g_{1}^{\pm }(x,x').
 \label{1-521}
 \end{equation}
As well from~(\ref{1-21}) we get the representations
\begin{align}
G_{1}^{}(x,x',& k)=  \notag \\
=& \frac{1}{2\pi }\int\limits_{k_{\Im }'=k_{\Im }}dk'\,
\Bigl[\theta (|k_{\Re }^{}|-|k_{\Re }'|)-\theta (x_{2}^{}-x_{2}')\Bigr]\Phi _{1}^{}(x,k')\Psi
_{1}^{}(x',k')+  \notag \\
& +c\,\theta (a-|k_{\Im }^{}|)\theta (x_{2}'-x_{2}^{})\Phi
_{1,-a}^{}(x)\Psi _{1,a}^{}(x'),  \label{1-53} \\
G_{1}^{\pm }(x,x',& k)=  \notag \\
=& \frac{1}{2\pi }\int\limits_{k_{\Im }'=k_{\Im }}dk_{\Re }'\,\Bigl[\theta \bigl((k_{\Im
}^{})_{}^{2}-a_{}^{2}-(k_{\Re }')_{}^{2}\bigr)-\theta (x_{2}^{}-x_{2}')\Bigr]\Phi
_{1}^{}(x,k')\Psi _{1}^{}(x',k')\mp  \notag \\
& \mp c\,\theta (a_{}^{2}-k_{\Im }^{2})\theta (\pm (x_{2}^{}-x_{2}'))\Phi _{1,-a}^{}(x)\Psi
_{1,a}^{}(x').  \label{1-54}
\end{align}
The first of these equalities shows that the cut of the resolvent at $
q_{2}=a^{2}$, $|q_{1}|<a$ is not inherited by $G_{1}(k)$ (in contrast to the case of the
nonstationary Schr\"{o}dinger equation, as mentioned in the Introduction) and that $G_{1}(k)$
is discontinuous only at the points $k=\pm ia$. Its behavior in the neighborhoods of these
points follows from~(\ref{1-25}) and reads as
 \begin{equation}
 G_{1}^{}(k)=g_{1}^{-}-\frac{c}{\pi }\left\{ \arccot \frac{a-|k_{\Im
 }^{}|}{|k_{\Re }^{}|}\right\} \Phi _{1,-a}^{}\otimes \Psi _{1,a}^{}+o(1),\quad k\sim \pm ia.
 \label{1-55}
 \end{equation}

Also the Green's functions $G_{1}^{\pm }(k)$ are discontinuous only at $k_\Im=\pm a$ and one
gets thanks to~(\ref{1-25}) that for $k\sim ia$ or $k\sim -ia$
 \begin{equation}
 G_{1}^{\pm }(k)=g_{1}^{-}-c\frac{1\pm \theta (a-|k_{\Im }^{}|)}{2}\Phi _{1,-a}^{}\otimes
 \Psi_{1,a}^{}+o(1).  \label{1-57}
 \end{equation}
Notice that these functions $G_{1}^{\pm }(k)$ coincide when $|k_{\Im }|>a$ and are independent
of $k_{\Im }$ (and then of $k$ by~(\ref{1-52})) when $|k_{\Im }|<a$. On the borders of these
strips they have the discontinuity
 \begin{equation}
 G_{1}^{\pm }(k)\Bigr|_{\text{outside}}^{}-G_{1}^{\pm }(k)\Bigr|_{\text{inside}}^{}=\pm
 \frac{c}{2}\Phi _{1,-a}^{}\otimes \Psi _{1,a}^{}. \label{1-56}
 \end{equation}
Taking into account the discontinuous behavior of the Green's functions we see that equalities
of the type $G_{1}^{\pm }(ia)=G_{1}^{}(ia)$ and $G_{1}^{\pm }(-ia)=G_{1}^{}(-ia)$ have no
meaning in our case. Thanks to~(\ref{1-53}) and~(\ref{1-57}) we have only that
 \begin{equation}
 \lim_{|k_{\Im }^{}|\rightarrow a-0}\lim_{k_{\Re }^{}\rightarrow 0}G_{1}^{}(k)=g_{1}^{-},
 \label{1-58}
 \end{equation}
where the limiting procedure must be performed in such way that
 $|k_{\Re}|/(a-|k_{\Im}|)\rightarrow +0$.

In order to complete the study of the Green's functions we mention that $G_{1}(k)$ obeys the
standard equalities
\begin{align}
& \lim_{k\rightarrow \infty }(-2ik)\frac{\partial }{\partial x_{1}^{}}
\,e^{i\ell (k)(x-x')}G_{1}^{}(x,x',k)=\delta
(x-x'),  \label{1-60} \\
& \frac{\partial G_{1}^{}(x,x',k)}{\partial \bar{k}}=\frac{\sgn
k_{\Re }^{}}{2\pi }\,\Phi _{1}^{}(x,-\bar{k})\Psi _{1}^{}(x',-
\bar{k}).  \label{1-61}
\end{align}
The first of them follows either from the differential equations~(\ref{1-50}),
or from the integral representation~(\ref{1-53}) and properties~(\ref{1-10}).
The second one also follows from~(\ref{1-50}), or it can be derived from~(\ref{1-47})
by means of~(\ref{1-44}) and~(\ref{1-45}). This equality must be understood in the sense of
distributions and we see that the
discontinuity of $G_{1}(k)$ at points $k=\pm ia$ leads (by~(\ref{1-7}),~(\ref{1-8}))
to the pole singularities of the r.h.s.\ at these points. In view of (\ref{1-61})
in what follows we refer to $G_{1}(k)$ as the Green's function of the Jost
solutions.

\section{Resolvent of the perturbed $L$-operator}

Now we consider the general case of the operator~(\ref{1}) with potential given
in~(\ref{potential}), where $u_{2}(x)$ is a real function of two space variables, smooth and
rapidly decaying at space infinity. The extended resolvent $M(q)$ is determined by (one of)
Eqs.~(\ref{17}) and we need to study its analyticity properties first. The increment
 $M(p;\mathbf{q}+\mathbf{s}) -M(p;\mathbf{q})$ of $M$ can be obtained from the Hilbert
identity~(\ref{H1}) where prime means the increment $\mathbf{s}$ of $\mathbf{q}$. We have
$M'-M=-M'(L_{1}'-L_{1}^{})M$ and, then, using~(\ref{1-3})
 \begin{equation}
 M_{}'-M=M_{}'L_{1}'(M_{1}'-M_{1}^{})L_{1}^{}M. \label{2-1}
 \end{equation}
Thus for the d-bar derivatives with respect to $\mathbf{q}_{j}$ we get
 \begin{equation}
 \bar{\partial}_{j}^{}M=(ML_{1}^{})(\bar{\partial}_{j}^{}M_{1}^{})(L_{1}^{}M),\quad j=1,2,
 \label{2-2}
 \end{equation}
in the region where $M_{1}$ is continuous, i.e.\ for $q_{2}\neq a^{2}$. In terms of the objects
introduced in~(\ref{11}), we obtain
 \begin{equation}
 \widehat{\bar{\partial}_{j}^{}M}=\widehat{M}\overleftarrow{\mathcal{L}}
 _{1}^{}(\widehat{\bar{\partial}_{j}^{}M_{1}})\overrightarrow{\mathcal{L}}
 _{1}^{}\widehat{M},\quad j=1,2,  \label{2-3}
 \end{equation}
where we used that $\widehat{L}_{1}(x,x';q)=\mathcal{L}_{1}\delta (x-x')$ and took into account
that when kernels with hat are considered the multiplication by $\mathcal{L}_{1}$ is no more
associative and it is necessary to use the arrows to indicate the correct order of operations
(cf.~(\ref{13})). Now, thanks to~(\ref{16}) and using~(\ref{1-44}) and~(\ref{1-45}) we get for
$q_{2}\neq a^{2}$
 \begin{align}
 \frac{\partial \widehat{M}(q)}{\partial q_{1}^{}}=& \frac{i}{\pi } \int\limits_{k_{\Im }=
 q_{1}}dk_{\Re }^{}\,\bar{k}\delta (\ell _{2\Im}^{}(k)-q_{2}^{})\,\Phi (k)\otimes \Psi (k),
 \label{2-4} \\
 \frac{\partial \widehat{M}(q)}{\partial q_{2}^{}}=& \frac{-1}{2\pi }\int\limits_{k_{\Im }=
 q_{1}}dk_{\Re }^{}\,\delta (\ell _{2\Im }^{}(k)-q_{2}^{})\,\Phi(k)\otimes \Psi (k),
 \label{2-5}
 \end{align}
where $\Phi (k)$ and $\Psi (k)$ are defined by
 \begin{equation}
 \Phi (k)=G(k)\overleftarrow{\mathcal{L}}_{1}^{}\Phi _{1}^{}(k),\quad \Psi (k)=\Psi
 _{1}^{}(k)\overrightarrow{\mathcal{L}}_{1}^{}G(k), \label{2-6}
 \end{equation}
with
\begin{equation}
G(x,x',k)=\widehat{M}(x,x';q)\Bigr|_{q=\ell _{\Im }(k)}^{}.  \label{2-7}
\end{equation}
More explicitly, say, the first of equations~(\ref{2-6}) stands for
 $\Phi (x,k) =\int dx'\bigl( \mathcal{L}_{1}^{\text{d}}(x',\partial_{x'})\times$
 $G_{1}(x,x',k)\bigr)\Phi _{1}(x',k)$.
The function $G(k)$ with kernel $G(x,x',k)$ defined in~(\ref{2-6}) satisfies the differential
equations
 \begin{equation}
 \overrightarrow{\mathcal{L}}G(k)=G(k)\overleftarrow{\mathcal{L}}=I, \label{2-9}
 \end{equation}
which can be obtained as a direct reduction of~(\ref{13}). Therefore, $G(k)$ is a Green's
function. Since the reduction is the same used in~(\ref{1-47}) for getting $G_{1}(k)$ from
$\widehat{M}_{1}$ we derive from~(\ref{17}) that this Green's function obeys the integral
equations
\begin{equation}
G(k)=G_{1}^{}(k)+\,G_{1}^{}(k)U_{2}^{}G(k),\qquad G(k)=G_{1}^{}(k)+\,G(k)U_{2}^{}G_{1}^{}(k).
\label{2-8}
\end{equation}
Again, as in Sec.\ 3, thanks to~(\ref{2-7}) and~(\ref{2-4}),~(\ref{2-5}) we get the d-bar
derivative of the Green's function in the form
 \begin{equation}
 \frac{\partial G(k)}{\partial \overline{k}}=\frac{\sgn k_{\Re }^{}}{2\pi }\,
 \Phi (-\bar{k})\otimes \Psi (-\bar{k}),  \label{2-10}
 \end{equation}
where $\Phi (k)$ and $\Psi (k)$ are defined in~(\ref{2-6}). These objects due to their
definition and~(\ref{2-8}) obey the integral equations
 \begin{equation}
 \Phi (k)=\Phi _{1}^{}(k)+\,G_{1}^{}(k)U_{2}^{}\Phi (k),\quad \Psi (k)=\Psi _{1}^{}(k)+\Psi
 (k)\,U_{2}^{}G_{1}^{}(k),  \label{2-11}
 \end{equation}
where again the first equation more explicitly reads as
 $\Phi (x,k) =\Phi _{1}^{}(x,k)+\int dx'G_{1}(x,x',k)\times$ $u_{2}(x')\Phi (x',k)$.
It is clear that the differential equations
\begin{equation}
\overrightarrow{\mathcal{L}}\Phi (k)=0,\qquad \Psi (k)\overleftarrow{
\mathcal{L}}=0,  \label{2-12}
\end{equation}
hold and, therefore, we can consider $\Phi (x,k)$ and $\Psi (x,k)$ as the generalization of the
Jost solutions to the case where the perturbation $u_{2}(x)$ is different from zero. Let us
mention that thanks to these definitions we succeeded to avoid the indeterminacy in the
definition of the Jost solutions discussed in the Introduction. Below we study the properties
of the Green's function and the Jost solutions in more details and discuss the singular
structure of the terms involved in~(\ref{2-10}). Now let us mention the following standard
properties
 \begin{align}
 & -2i\lim_{k\rightarrow \infty }k\partial _{x_{1}}^{}\left( e_{}^{i\ell(k)(x-x')}G(x,x',k)
 \right)=\delta (x-x'),  \label{2-13} \\
 & \overline{G(k)}=G(-\bar{k})=G(k),  \label{2-14} \\
 & \overline{\Phi (x,k)}=\Phi (x,-\overline{k}),\qquad \overline{\Psi (x,k)}
 =\Psi (x,-\overline{k}),  \label{2-15}
 \end{align}
that can be obtained by means of the integral equations~(\ref{2-8}) and properties~(\ref{1-9}),
(\ref{1-51}), and~(\ref{1-60}) for the Green's function $G_{1}(k)$.

Till now we studied the departure from analyticity of the resolvent in the case
 $q_{2}\neq a^{2}$. Since the resolvent $M_{1}(q)$ is discontinuous along the line
 $q_{2}=a^{2}$ (see~(\ref{1-46})), the integral equations~(\ref{17}) suggest that also $M(q)$
has a discontinuity. Let us denote the limiting values on the two edges of the line by
 \begin{equation}
 M_{}^{\pm }(q)=M(q)\Bigr|_{q_{2}=a^{2}\pm 0}^{}.  \label{2-16}
 \end{equation}
Then from the Hilbert identity~(\ref{2-1}) we derive that
 \begin{equation}
 M_{}^{+}(q)-M_{}^{-}(q)=M_{}^{\pm
 }(q)L_{1}^{}(q)(M_{1}^{+}(q)-M_{1}^{-}(q))L_{1}^{}(q)M_{}^{\mp }(q),\quad q_{2}^{}=a^{2},
 \label{2-17}
 \end{equation}
where the l.h.s.\ is independent of the choice of the sign in the r.h.s.\ In analogy
with~(\ref{1-48}) we introduce the two Green's functions
 \begin{equation}
 G_{}^{\pm }(x,x';k)=\widehat{M}(x,x';q)\Bigr|_{q_{1}=k_{\Im ,}\,q_{2}=a^{2}\pm 0}^{}
 \label{2-18}
 \end{equation}
and rewrite~(\ref{2-17}) in these terms as
 $G_{}^{+}(k)-G_{}^{-}(k)=(G_{}^{\pm }(k)\overleftarrow{\mathcal{L}}
 _{1}^{})(G_{1}^{+}(k)-G_{1}^{-}(k))\times\linebreak(\overrightarrow{\mathcal{L}}
 _{1}^{}G_{}^{\mp }(k))$.
Then by~(\ref{1-46}) and~(\ref{1-48}) we get
 \begin{equation}
 G_{}^{+}(k)-G_{}^{-}(k)=-c\theta (a_{}^{2}-k_{\Im }^{2})\Phi _{}^{\pm }(k)\otimes
 \Psi _{}^{\mp}(k),  \label{2-20}
 \end{equation}
where the new solutions (cf.~(\ref{2-6}))were introduced:
 \begin{equation}
 \Phi _{}^{\pm }(k)=G_{}^{\pm }(k)\overleftarrow{\mathcal{L}} _{1}^{}\Phi _{1,-a}^{},\qquad
 \Psi _{}^{\pm }(k)=\Psi _{1,a}^{}\overrightarrow{\mathcal{L}}_{1}^{}G_{}^{\pm }(k).
 \label{2-21}
 \end{equation}
Following properties of $G_{1}^{\pm }(k)$ it is easy to show that these Green's functions obey
the following differential and integral equations and reality condition
 \begin{align}
 & \overrightarrow{\mathcal{L}}G_{}^{\pm }(k)=G_{}^{\pm }(k)
 \overleftarrow{\mathcal{L}}=I,  \label{2-22} \\
 & G_{}^{\pm }(k)=G_{1}^{\pm }(k)+\,G_{1}^{\pm }(k)U_{2}^{}G_{}^{\pm }(k),\qquad G_{}^{\pm}(k)=
 G_{1}^{\pm }(k)+\,G_{}^{\pm}(k)U_{2}^{}G_{1}^{\pm }(k),  \label{2-23} \\
 & \overline{G_{}^{\pm }(k)}=G_{}^{\pm }(k).  \label{2-24}
 \end{align}
By definition they are independent of $k_{\Re }$ and by the corresponding properties of
$G_{1}^{\pm }(k)$ we have that $G^{+}(k)=G^{-}(k)$ when $|k_{\Im }|>a$ and they are independent
of $k_{\Im }$ when $|k_{\Im }|<a$. By~(\ref{2-21}) and~(\ref{2-22}) we get that $\Phi ^{\pm
}(x,k)$ and $\Psi ^{\pm }(k)$ are solutions of the heat equation with
potential~(\ref{potential}),
 \begin{equation}
 \overrightarrow{\mathcal{L}}\Phi _{}^{\pm }(k)=0,\qquad
 \Psi_{}^{\pm}(k)\overleftarrow{\mathcal{L}}=0.  \label{2-25}
 \end{equation}
Integral equations for these solutions follow by applying operations~(\ref{2-21}) to the
equations~(\ref{2-23})
 \begin{equation}
 \Phi _{}^{\pm }=\Phi _{1,-a}^{}+\,G_{1}^{\pm }U_{2}^{}\Phi _{}^{\pm },\qquad
 \Psi_{}^{\pm}=\Psi _{1,a}^{}+\Psi _{}^{\pm }\,U_{2}^{}G_{1}^{\pm }.  \label{2-26}
 \end{equation}
Let us also mention that thanks to~(\ref{2-24}) these solutions are real and are independent of
$k$ inside the strip $|k_{\Im }|<a$, due to the corresponding property of $G^{\pm }(k)$
and~(\ref{2-21}). Since in the following we use intensively the Green's functions and these
solutions inside the strip it is convenient to introduce the following specific notations
 \begin{equation}
 g_{}^{\pm}(x,x')=G_{}^{\pm }(x,x',k)\Bigr|_{|k_{\Im }|<a}^{},\label{2-261}
 \end{equation}
and also
 \begin{equation}
 \phi _{}^{\pm }(x)=\Phi _{}^{\pm }(x,k)\Bigr|_{|k_{\Im}|<a}^{},\qquad
 \psi _{}^{\pm }(x)=\Psi _{}^{\pm }(x,k)\Bigr|_{|k_{\Im }|<a}^{}.\label{2-262}
 \end{equation}

Equality~(\ref{2-20}) enables us to find relations between solutions~(\ref{2-21}).
Let $|k_{\Im }|<a$, then applying, say, $\overleftarrow{\mathcal{L}}_{1}^{}
\Phi _{1,-a}^{}$ to this equality from the right and using~(\ref{2-21}) we derive that
\begin{equation}
(1+\lambda )\phi _{}^{+}=\phi _{}^{-},  \label{2-27}
\end{equation}
where
\begin{equation}
\lambda =c(\Psi _{1,a}^{}\overrightarrow{\mathcal{L}}_{1}^{}g_{}^{-}
\overleftarrow{\mathcal{L}}_{1}^{}\Phi _{1,-a}^{}).  \label{2-28}
\end{equation}
Explicitly $\lambda =c\int dx\int dx'\Psi _{1,a}^{}(x)(\mathcal{L}_{1}^{}(x,\partial _{x})
\mathcal{L}_{1}^{\text{d}}(x',\partial
_{x'})g_{}^{-}(x,x')\Phi _{1,-a}^{}(x')$. By (\ref{2-24}) this constant is real
and thanks to~(\ref{2-21}) it is also equal to $\lambda
=c(\Psi _{1,a}\overrightarrow{\mathcal{L}}_{1}\phi ^{-})=c(\psi
^{-}\overleftarrow{\mathcal{L}}_{1}\Phi _{1,-a})$. Inserting
here $\mathcal{L}_{1}=\mathcal{L}+U_{2}$ we get by~(\ref{2-22}) and~(\ref{2-25}) that
 \begin{equation}
 \lambda =c\{(\Psi _{1,a}^{}U_{2}^{}\Phi _{1,-a}^{})+(\Psi _{1,a}^{}U_{2}^{}g_{}^{-}U_{2}^{}
 \Phi_{1,-a}^{})\},  \label{2-29}
 \end{equation}
or $\lambda =c(\Psi _{1,a}U_{2}\phi ^{-})=c(\psi ^{-}U_{2}\Phi _{1,-a})$, where we also used
$\overrightarrow{\mathcal{L}}_{1}\Phi _{1,-a}=0$ and
 $\Psi_{1,a}\overleftarrow{\mathcal{L}}_{1}=0$ that follows from~(\ref{1-6}) and~(\ref{1-12}).
Next, applying to~(\ref{2-27}) $\Psi _{1,a}^{}\overrightarrow{\mathcal{L}}_{1}^{}$ from the
left and again by~(\ref{2-21}) we get
 \begin{equation}
 (1+\lambda )[1-c(\Psi _{1,a}^{}\overrightarrow{\mathcal{L}}
 _{1}^{}g_{}^{+}\overleftarrow{\mathcal{L}}_{1}^{}\Phi _{1,-a}^{})]=1, \label{2-30}
 \end{equation}
where a new constant
 $(\Psi_{1,a}\overrightarrow{\mathcal{L}}_{1}g^{+}\overleftarrow{\mathcal{L}}_{1}\Phi _{1,-a})
 =(\Psi_{1,a}U_{2}\Phi _{1,-a})+(\Psi _{1,a}U_{2}g^{+}U_{2}\Phi _{1,-a})$ (cf. (\ref{2-29}))
appeared. Since we chose $u_{2}$ to be rapidly decaying at infinity all terms must be finite.
Then $1+\lambda \neq 0$ and, more precisely, taking into account that for $u_{2}\rightarrow 0$
also $\lambda \rightarrow 0$ we have that
 \begin{equation}
 1+\lambda >0.  \label{2-31}
 \end{equation}

Summarizing, we get the following relations:
 \begin{align}
 & c(\Psi _{1,a}^{}\overrightarrow{\mathcal{L}}_{1}^{}g_{}^{+}
 \overleftarrow{\mathcal{L}}_{1}^{}\Phi _{1,-a}^{})=\frac{\lambda }{1+\lambda },\label{2-32} \\
 & \phi _{}^{+}=\frac{\phi _{}^{-}}{1+\lambda },\qquad \psi _{}^{+}=
 \frac{\psi _{}^{-}}{1+\lambda },  \label{2-33} \\
 & G_{}^{+}(k)=G_{}^{-}(k)-\frac{c\theta (a-|k_{\Im }^{}|)}{1+\lambda }
 \phi _{}^{-}\otimes \psi _{}^{-}.  \label{2-34}
 \end{align}
Here~(\ref{2-32}) is just~(\ref{2-30}), the first equality in~(\ref{2-33}) is~(\ref{2-28}) and
the second equality is derived by analogy, and~(\ref{2-34}) follows from~(\ref{2-20}) thanks
to~(\ref{2-33}). In their turn~(\ref{2-32}) and~(\ref{2-33}) follow from~(\ref{2-34}) thanks
to~(\ref{2-21}) and~(\ref{2-27}).

We have shown in~(\ref{1-56}) that the Green's functions $G_{1}^{\pm }(k)$ are discontinuous at
$k_{\Im }=a$ and $k_{\Im }=-a$. By~(\ref{2-23}) we deduce that $G_{}^{\pm }(k)$ have the same
behavior. In order to study this discontinuity we use, as above, the Hilbert
identity~(\ref{2-1}) where $M=M(q)$, $M'=M(q')$, etc. We choose $q_{2}=q_{2}'=a^{2}\pm 0 $,
$q_{1}=a-\varepsilon $, $q_{1}'=a+\varepsilon $, and in the limit $\varepsilon \rightarrow +0$
we use the hat notation~(\ref{11}) and definitions~(\ref{1-48}) and~(\ref{2-18}) of the Green's
functions. Then we get
 $G^{\pm }(i(a+0))-g^{\pm }=G^{\pm }(i(a+0))\overleftarrow{\mathcal{L}}_{1}
 (G_{1}^{\pm }(i(a+0))-g_{1}^{\pm })\overrightarrow{\mathcal{L}}_{1}g^{\pm }$,
where again~(\ref{1-521}) and~(\ref{2-261}) where used. Now by~(\ref{1-56}) for the
discontinuity of the unperturbed Green's functions we obtain
 \begin{equation}
 G_{}^{\pm }(i(a+0))-g_{}^{\pm }=\pm \frac{c}{2}\Phi _{}^{\pm }(i(a+0))\otimes \psi _{}^{\pm }
 \label{2-35}
 \end{equation}
where notations~(\ref{2-21}),~(\ref{2-261}) and~(\ref{2-262}) were used.
Applying $\overleftarrow{\mathcal{L}}_{1}\Phi _{1,-a}$ from the right and
 $\Psi _{1,a}\overrightarrow{\mathcal{L}}_{1}$ from the left in analogy with the derivation
of~(\ref{2-34}) we get by~(\ref{2-27}) that
 $c(\Psi _{1,a}\overrightarrow{\mathcal{L}}_{1}G^{\pm }(i(a+0))\overleftarrow{\mathcal{L}}_{1}
 \Phi _{1,-a})=2\lambda (2+\lambda)^{-1}$, that is finite
due to~(\ref{2-31}). Then omitting details we derive the equalities
 $G^{\pm }(i(a+0))=G^{\pm }(-i(a+0))=g^{-}-c(2+\lambda)^{-1}\phi _{}^{-}\otimes \psi _{}^{-}$,
that, say, for the bottom sign can also be rewritten in the form
 \begin{equation}
 G_{}^{-}(k)=g_{}^{-}-\frac{c\theta (|k_{\Im }^{}|-a)}{2+\lambda }\phi _{}^{-}\otimes
 \psi _{}^{-}+o(1),\quad k\sim \pm ia, \label{2-36}
 \end{equation}
where we took~(\ref{2-261}) into account.

\section{Properties of the Jost solutions and Inverse problem}

In this section we complete the investigation of the properties of the Jost solutions by
describing their behavior at the points $k=\pm ia$. Formulae~(\ref{2-6}) suggest to study first
the behavior of the Green's function $G(k)$. We expect that it is ill defined at these points,
so in order to describe this behavior we compare $G(k)$ with some well defined Green's
function, say, $g^{-}$. For this aim, as we have already shown, relations of the
type~(\ref{2-20}) can be very useful. In order to derive them we start again from the Hilbert
identity~(\ref{2-1}) where $M'=M(q')$ and $M=M(q)$ and we choose
 $q'=\ell _{\Im }(k)$, $q_{1}=k_{\Im }$, $q_{2}=a^{2}-0$ (see~(\ref{l}), (\ref{2-7})
and~(\ref{2-16})). Then, passing to the objects with hats by~(\ref{11}), recalling
definitions~(\ref{2-7}),~(\ref{2-18}) and keeping only the leading term in the neighborhood of
$k\sim \pm ia$ we get
 \begin{equation*}
 G(k)-G_{}^{-}(k)=G(k)\overleftarrow{\mathcal{L}}_{1}^{}(G_{1}^{}(k)-G_{1}^{-}(k))
 \overrightarrow{\mathcal{L}}_{1}^{}G_{}^{-}(k)+o(1),\quad k\sim \pm ia.
 \end{equation*}
Inserting the explicit singular behaviors of $G_{1}(k)$, $G_{1}^{-}(k)$ and $G^{-}(k)$ at
$k=\pm ia$ given in~(\ref{1-55}),~(\ref{1-57}) and~(\ref{2-36}), we have
 \begin{align*}
 G(k)& -g_{}^{-}=-\frac{c\theta (|k_{\Im }^{}|-a)}{2+\lambda }\phi_{}^{-}\otimes \psi_{}^{-}+\\
 & +c\left( -\frac{1}{\pi }\arccot\frac{a-|k_{\Im }^{}|}{|k_{\Re}^{}|}+
 \frac{\theta (|k_{\Im }^{}|-a)}{2}\right) G(k)\overleftarrow{\mathcal{L}}_{1}^{}
 \Phi _{1,-a}^{}\otimes \Psi _{}^{-}(k)+o(1),
 \end{align*}
where in the last multiplier the definition of $\Psi ^{-}(k)$ in~(\ref{2-21}) was used. Again
by~(\ref{2-21}) and~(\ref{2-36})
 $\Psi ^{-}(k)=\frac{2+\lambda \theta (a-|k_{\Im }|)}{2+\lambda }\psi ^{-}$, where as always
 $\psi ^{-}$ denotes $\Psi ^{-}(k)$ for $|k_{\Im }|<a$ by~(\ref{2-262}). Then
 \begin{align}
 G(k)-g_{}^{-}=& \,\Biggl\{-\frac{c\theta (|k_{\Im }^{}|-a)}{2+\lambda }\phi _{}^{-}+
 c\left( -\frac{1}{\pi }\arccot\frac{a-|k_{\Im }^{}|}{|k_{\Re }^{}|}+
 \frac{\theta (|k_{\Im }^{}|-a)}{2}\right) \times\notag \\
 & \times \frac{2+\lambda \theta (a-|k_{\Im }^{}|)}{2+\lambda }G(k)
 \overleftarrow{\mathcal{L}}_{1}^{}\Phi _{1,-a}^{}\Biggr\}\otimes \psi ^{-}+o(1). \label{2-37}
 \end{align}
Thus in order to get the behavior of $G(k)$ we need to find that of
 $G(k)\overleftarrow{\mathcal{L}}_{1}\Phi _{1,-a}$, which follows by applying to~(\ref{2-37})
operation $\overleftarrow{\mathcal{L}}_{1}\Phi _{1,-a}$ from the right and using
again~(\ref{2-21}),~(\ref{2-262}) and~(\ref{2-28}). Then
 \begin{equation}
 G(k)\overleftarrow{\mathcal{L}}_{1}^{}\Phi _{1,-a}^{}=\frac{\pi \phi _{}^{-}}{A(k)}+o(1),\quad
 k\sim\pm ia,  \label{2-38}
 \end{equation}
where we denoted for shortness
 \begin{equation}
 A(k)=\pi +\lambda \arccot\frac{a-|k_{\Im }^{}|}{|k_{\Re }^{}|}. \label{Ak}
 \end{equation}
This function is real, positive thanks to~(\ref{2-31}) and discontinuous at $k=\pm ia$. Now
inserting~(\ref{2-38}) in~(\ref{2-37}) we derive finally that
 \begin{equation}
 G(k)=g_{}^{-}-\frac{c}{A(k)}\left(\arccot\frac{a-|k_{\Im }^{}|}{|k_{\Re }^{}|}\right)
 \,\phi _{}^{-}\otimes \psi _{}^{-}+o(1),\quad k\sim \pm ia.  \label{2-39}
 \end{equation}
Applying to~(\ref{2-39}) from the left the operation
 $\Psi _{1,a}^{}\overrightarrow{\mathcal{L}}_{1}^{}$ and recalling the definitions~(\ref{2-21})
and~(\ref{2-262}) we derive
 \begin{equation}
 \Psi _{1,a}^{}\overrightarrow{\mathcal{L}}_{1}^{}G(k)=\frac{\pi \psi _{}^{-}}{A(k)}+o(1),\quad
 k\sim \pm ia,  \label{2-40}
 \end{equation}
and by~(\ref{2-28}) also
 \begin{equation}
 c\Psi _{1,a}^{}\overrightarrow{\mathcal{L}}_{1}^{}G(k)\overleftarrow{\mathcal{L}}_{1}^{}
 \Phi _{1,-a}^{}=\frac{\pi \lambda }{A(k)}+o(1),\quad k\sim\pm ia.
 \label{2-41}
 \end{equation}
Correspondingly, we get for the behavior of the Jost solutions in the neighborhood of
 $k=\pm ia$, thanks to~(\ref{1-7}) and~(\ref{2-38})
\begin{align}
\Phi (k)& =\frac{i\pi c\phi _{}^{-}}{A(k)(k-ia)}+O(1)\,,\quad k\sim ia,
\label{Phi+} \\
\Phi (k)& =\frac{\pi \,\phi _{}^{-}}{A(k)}+o(1),\quad k\sim -ia \label{Phi-}
\end{align}
and analogous relations for $\Psi (k)$.

Now we are ready to consider the d-bar derivative in the sense of distributions of the Jost
solution, say, $\Phi (k)$. Let first $k\neq \pm ia$. Then we use~(\ref{1-7}),~(\ref{2-6}),
and~(\ref{2-10}) to derive
 \begin{equation}
 \frac{\partial \Phi (k)}{\partial \bar{k}}=\Phi (-\bar{k})r(k),\quad k\neq \pm ia,
 \label{dbarPhi'}
 \end{equation}
where the Spectral data are defined as follows
 \begin{equation}
 r(k)=\frac{\sgn k_{\Re }^{}}{2\pi }\,\Bigl(\Psi _{1}^{}(-\bar{k})
 \overrightarrow{\mathcal{L}}_{1}^{}G(k)\overleftarrow{\mathcal{L}} _{1}^{}
\Phi _{1}^{}(k)\Bigr).  \label{rkpert}
 \end{equation}
Thanks to~(\ref{1-7}),~(\ref{1-8}) and~(\ref{2-41}) we get the singular behavior of these
spectral data in the form
 \begin{equation}
 r(k)=\frac{i\lambda \sgn k_{\Re }^{}}{2(k_{\Re }^{}+i|k_{\Im }^{}|-ia)A(k)}+o(1),\quad k\sim
 \pm ia,  \label{rksing}
 \end{equation}
i.e.\ in both points it has a pole singularity multiplied by the discontinuous function $A(k)$.
Taking into account that the singular behavior of $\Phi (-\bar{k})$ is given by the denominator
$\overline{(k-ia)}A(k)$ at point $k=ia$ and by $A(k)$ at point $k=-ia$ we see that the r.h.s.\
in~(\ref{dbarPhi'}) is integrable at the latter point but it has a singularity
 $\sgn k_{\Re}^{}|k-ia|^{-2}A(k)^{-2}$ at point $k=ia$, that is not integrable. On the other
side, $\Phi (k)$ is locally integrable for any $k$ so $\Phi (x,k)e^{i\ell (k)x}$ is a Schwartz
distribution with respect to $k$. Thus its d-bar derivative in the sense of distributions
exists and can be defined in the standard way. Let $f(k)$ be a test function that properly
decays at infinity (we are not interested in the exponential growth due to the multiplier
$e^{-i\ell (k)x}$ now). Then the d-bar derivative of $\Phi (k)$ is defined as
 \begin{equation*}
 \int d_{}^{2}k\,\frac{\partial \Phi (k)}{\partial \bar{k}}f(k)=-\int
 d_{}^{2}k\,\Phi (k)\frac{\partial f(k)}{\partial \bar{k}}
 =-\lim_{\varepsilon \rightarrow 0}\int\limits_{|k\pm ia|>\varepsilon }d_{}^{2}k\,\Phi
 (k)\frac{\partial f(k)}{\partial \bar{k}},
 \end{equation*}
where in the last equality we again used the property of local integrability of $\Phi (k)$.
Integrating by parts for $\varepsilon >0$ we can use~(\ref{dbarPhi'}) and we have
 \begin{align*}
 -\lim_{\varepsilon \rightarrow 0}\int\limits_{|k\pm ia|>\varepsilon
 }d_{}^{2}k\,& \,\Phi (k)\frac{\partial f(k)}{\partial \bar{k}}= \\
 & =\frac{f(ia)}{2i}\lim_{\varepsilon \rightarrow 0}\oint\limits_{|k-ia|=\varepsilon }dk\,\Phi
 (k)+\lim_{\varepsilon
 \rightarrow 0}\int\limits_{|k-ia|>\varepsilon }d_{}^{2}k\,\Phi (-\overline{ k})r(k)f(k),
 \end{align*}
where we omitted the term $\oint_{|k+ia|>\varepsilon }$ since thanks to~(\ref{Phi-}) it gives
zero in the limit $\varepsilon \rightarrow 0$. Thanks to~(\ref{Phi+}) and~(\ref{Phi-}) both
limits in the r.h.s.\ exist. To be more precise let us introduce the distribution
 \begin{equation}
 \text{p.v.}\int d_{}^{2}k\frac{\sgn k_{\Re }^{}f(k)}{|k-ia|^{2}A(k)^{2}}
 =\lim_{\varepsilon \rightarrow 0}\int\limits_{|k-ia|>\varepsilon }d_{}^{2}k
 \frac{\sgn k_{\Re }^{}f(k)}{|k-ia|^{2}A(k)^{2}}.  \label{distr1}
 \end{equation}
Notice the presence in the numerator of $\sgn k_{\Re }^{}$ that guaranties existence of the
limit. We used the principal value (p.v.) notation in analogy with the one-dimensional case. It
can be checked directly that
 \begin{align}
 \text{p.v.}\int d_{}^{2}k\frac{\sgn k_{\Re }^{}f(k)}{|k-ia|^{2}A(k)^{2}} & =\int
 d_{}^{2}k\,\frac{\sgn k_{\Re }^{}}{|k-ia|^{2}A(k)^{2}}
 \,[f(k)-\theta (\delta -\left| k-ia\right| )f(ia)]=  \notag \\
 & =\frac{1}{2}\int d_{}^{2}k\,\sgn k_{\Re }^{}\,\frac{f(k)-f(-\bar{k})}{|k-ia|^{2}A(k)^{2}},
 \label{distr2}
 \end{align}
where $\delta $ is some real positive parameter and the second term in~(\ref{distr2}) is
independent on the choice of $\delta $. In the case where a distribution has singularities of
this form at some finite number of points $a_{1}$, $a_{2}$, etc., we use the same notation for
the integral assuming that either the cutoff procedure in~(\ref{distr1}) or the subtraction
procedure in~(\ref{distr2}) is performed at each point. Of course, the parameters
 $\varepsilon_{j}$ and $\delta _{j}$ must be chosen in such way that corresponding discs do
not overlap.

Let us denote
 \begin{equation}
 \Phi _{a}^{}=-\frac{1}{2\pi }\lim_{\varepsilon \rightarrow 0}\oint\limits_{|k-ia|=\varepsilon
 }dk\,\Phi (k),  \label{regul2}
 \end{equation}
so that $i\Phi _{a}^{}$ can be considered as an extension of the definition of residuum to the
case in which the pole singularity is multiplied by a function discontinuous at the same point.
Thanks to~(\ref{Phi+}) we get that this limit also exists and equals
 \begin{equation}
 \Phi _{a}^{}=c\frac{\log (1+\lambda )}{\lambda }\,\phi _{}^{-}. \label{regul+}
 \end{equation}
Thus, summarizing all above definitions we get that
 \begin{equation}
 \frac{\partial \Phi (k)}{\partial \bar{k}}=\Phi (-\bar{k})r(k)+i\pi \Phi _{a}^{}\delta (k-ia),
 \label{dbarPhi}
 \end{equation}
where $\Phi (x,-\bar{k})r(k)$ is now a distribution in $k$ defined by the p.v.\ prescription
given above. By~(\ref{Phi+}),~(\ref{Phi-}), and~(\ref{rksing}) it is integrable at $k=-ia$, but
it behaves as $|k-ia|^{-2}A^{-2}(k)$ in the neighborhood of the point $k=ia$.

Equation~(\ref{dbarPhi}) supplies us with the first equation of the Inverse problem. In order
to close it we need the analog of the first relation in~(\ref{1-14}), where it is stated that
the residuum of the function is proportional to its value in the conjugated point. But in our
case $\Phi(k)$ is discontinuous at point $k=-ia$, so again some modification of the notion of
``value'' at this point must be given. Following the procedure used in~(\ref{regul2}) we can
define it as
 \begin{equation}
 \Phi _{-a}^{}=\frac{1}{2\pi i}\lim_{\varepsilon \rightarrow 0}\oint\limits_{|k+ia|=\varepsilon}
 \frac{dk}{k+ia}\,\Phi (k). \label{regul3}
 \end{equation}
Thanks to~(\ref{Phi-}) this limit also exists and equals
 \begin{equation}
 \Phi _{-a}^{}=\frac{\log (1+\lambda )}{\lambda }\,\phi _{}^{-}, \label{regul4}
 \end{equation}
so that by~(\ref{regul+}) we have
 \begin{equation}
 \Phi _{a}^{}=c\Phi _{-a}^{},  \label{discrPhi}
 \end{equation}
that shows that the parameter $c$ is not modified by the perturbation. This equality closes the
formulation of the Inverse problem~(\ref{dbarPhi}). Finally, taking into account the asymptotic
behavior of $\Phi (x,k)$ and~(\ref{regul3}) and~(\ref{discrPhi}) we can formulate the Inverse
problem as the following system of integral equations:
 \begin{align}
 \Phi (x,k)& =e_{}^{-i\ell (k)x}+  \notag \\
 & +\frac{1}{\pi }\text{p.v.}\int \frac{d_{}^{2}k'}{k-k'}\,e_{}^{i(\ell (k')-\ell (k))x}
 \Phi (x,-\overline{k'})r(k')+i\frac{e_{}^{i(\ell (ia)-\ell (k))x}}{k-ia}\,\Phi_{a}^{}(x),
 \label{2-42} \\
 \frac{1}{c}\Phi _{a}^{}(x)& =e_{}^{-i\ell (-ia)x}-  \notag \\
 & -\frac{1}{\pi }\text{p.v.}\int d_{}^{2}k\,\frac{\Phi (x,-\overline{k})r(k)}
 {k+ia}e_{}^{i(\ell (k)-\ell (-ia))x}+\frac{e_{}^{i(\ell (ia)-\ell (-ia))x}}{2a}\,
 \Phi_{a}^{}(x).  \label{2-43}
 \end{align}
The integrands in the r.h.s.\ of these two equations are not locally integrable, respectively,
the first at $k=ia$ and the second at $k=\pm ia$. Correspondingly, their integrals are
regularized by means of the principal value prescription, as in~(\ref{distr1})
or~(\ref{distr2}), at $k=ia$ and at $k=\pm ia$.

The potential is reconstructed by means of
 \begin{equation}
 u(x)=-\frac{2i}{\pi }\text{p.v.}\int d_{}^{2}k\,\frac{\partial}{\partial x_{1}}\!\left( \,e_{}^{i\ell (k)x}\Phi
 (x,-\overline{k})r(k)\right)+2\frac{\partial}{\partial x_{1}}\!\left( e_{}^{i\ell (ia)x}\,\Phi _{a}^{}(x)\right).  \label{2-44}
 \end{equation}

\section{Conclusion}

In this article on the basis of the resolvent approach we gave a detailed presentation of an
extension of the inverse scattering theory for the heat operator to the case where the
potential~(\ref{potential}) is a perturbation of the 1-dimensional one soliton potential
$u_1(x_1)$~(\ref{1-5}) by means of a smooth, decaying at infinity function $u_2(x)$ of two
space variables. To our knowledge this is the first time that inverse scattering theory is
applied to a non-scattering situation, i.e.\ a situation with an infinite obstacle. As a
result of our investigation we proved that under such a perturbation the Jost solutions get
specific singularities~(\ref{Phi+}) and~(\ref{Phi-}) on the complex plane of the spectral
parameter $k$. We demonstrated that the d-bar problem~(\ref{dbarPhi}),~(\ref{discrPhi}), while
looking familiar for a potential whose spectrum has a discrete and continuous part, needs a
substantially modified approach due to the singularity structure of the spectral data given
in~(\ref{rksing}). It was necessary to establish the meaning in the sense of distributions of
all terms involved in this problem, in order to be able to formulate the Inverse problem as
the system of integral equations (\ref{2-42}),~(\ref{2-43}). It is easy to check that the
singular behavior of the spectral data and Jost solution as given in~(\ref{rksing})
and~(\ref{Phi+}),~(\ref{Phi-}) is compatible with this Inverse problem. On the other side, it
is necessary to prove that the potential $u(x)$ reconstructed by means of~(\ref{2-44}) is of
the type~(\ref{potential}). We plan to address this problem in a forthcoming work.

Another open problem is the application of these results to the KPII
equation~(\ref{KPII}) itself. In particular, investigation of the time asymptotics of solutions
with initial data of the type~(\ref{potential}) must be performed. Let us mention only that the
singular behavior~(\ref{rksing}) of the spectral data is preserved under
evolution~(\ref{KPII}). Indeed~\cite{BarYacoov}, the time dependence of the spectral data is given
as
 \begin{equation}
 r(k,t)=e_{}^{4i(k^3+\bar{k}^3)}.  \label{rkt}
 \end{equation}
Thus we get that
 \begin{equation}
 a={\rm const},\qquad \lambda={\rm const}  \label{alambdat}
 \end{equation}
also with respect to time.

In Sec.~3 we mentioned that the above construction can be easily generalized to the
case where the potential $u_1(x_1)$ is a pure $N$-soliton 1-dimensional potential. At the same time
our approach also admits straightforward generalization to the case where $u_1(x)$ is not a
function of one space variable but the result of application of the B\"{a}cklund transformation
to a generic background 2-dimensional potential $u_0(x)$ decaying on the $x$-plane. Then the
inverse problem is again given by Eqs.~(\ref{2-42}) and~(\ref{2-43}), where the spectral data
$r(k)$ are replaced with
 \begin{equation}
 r(k)+\frac{(k+ia)(\bar{k}+ia)}{({\bar{k}}-ia)(k -ia)} r^{}_0(k), \label{2rk}
 \end{equation}
where $r(k)$ is of the type~(\ref{rksing}) and $r_0(k)$ are the spectral data of the potential
$u_0(x)$ (see~\cite{KPII}).

The theory of the heat equation with respect to the nonstationary Schr\"{o}dinger equation is
in some respects simpler and in some other respects unexpectedly more difficult. As we have
shown, under perturbation the Jost solution get singularities more complicated than poles, but
this solution has no additional cut in the complex domain, in contrast with the
nonstationary Schr\"{o}dinger case as discovered in~\cite{FokasPogreb}. On the other side
the generalization of this scheme to the case of multi-ray structure of the potential $u(x)$ meets
with essential problems, first of all due to the fact that the resolvent (or Green's function)
of the heat equation even of a 2-soliton (generic) potential is unknown in the literature. This
problem also needs future development.

\noindent\textbf{Acknowledgment}

A.K.P. thanks his colleagues at the Department of Physics of the University of Lecce for kind
hospitality and S.V.Manakov and V.E.Zakharov for fruitful discussions.

\end{document}